\documentclass[runningheads]{llncs}
\usepackage[T1]{fontenc}
\usepackage{amsmath}
\usepackage{amsfonts}
\usepackage{graphicx}
\usepackage{orcidlink}

\usepackage{amsthm}
\usepackage{amsmath, amssymb, mathtools}
\usepackage{bm}
\usepackage{booktabs, tabularx}
\usepackage{siunitx}
\usepackage{tikz}
\usetikzlibrary{arrows.meta, positioning, shapes.misc, calc}
\usepackage{algorithm, algorithmic} % optional for pseudocode

\begin{document}
\title{Geometric Hyperscanning of Affect \\ under Active
Inference}
\author{Nicol\'as Hinrichs\inst{1, 2}\orcidlink{0000-0003-4969-9644} \and
Mahault Albarracin\inst{3, 4}\orcidlink{0000-0003-0916-4645} \and
Dimitris Bolis\inst{5}\orcidlink{0000-0001-9656-8685} \and
Yuyue Jiang\inst{6,7}\orcidlink{0009-0001-2896-7323} \and
Leonardo Christov-Moore\inst{8,9}\orcidlink{0000-0003-1589-5321}\and
Leonhard Schilbach\inst{10,11}\orcidlink{0000-0001-9656-8685}}
\authorrunning{Hinrichs, Albarracin, Bolis, et al.}
\institute{
Max Planck Institute for Human Cognitive and Brain Sciences, Leipzig, Germany \and Okinawa Institute of Science and Technology, Okinawa, Japan \and VERSES AI Research Lab, Los Angeles, USA \and Universit\'e du Qu\'ebec \`a Montr\'eal, Montr\'eal (Qu\'ebec), Canada \and Instituto Italiano di Tecnologia, Rovereto, Italy \and University of California, Santa Barbara, USA \and University of California, Irvine, USA \and Institute for Advanced Consciousness Studies, Santa Monica, USA \and SENSORIA Research, San Francisco, CA \and Ludwig Maximilians Universit\"at, Munich, Germany \and LVR-Klinikum D\"usseldorf / Heinrich Heine University, D\"usseldorf, Germany
\email{hinrichsn@cbs.mpg.de nicolas.hinrichs@oist.jp}}
\maketitle              % typeset the header of the contribution
\begin{abstract}Second-person neuroscience holds social cognition as embodied meaning co-regulation through reciprocal interaction, modeled here as coupled active inference with affect emerging as inference over identity-relevant surprise. Each agent maintains a self-model that tracks violations in its predictive coherence while recursively modeling the other. Valence is computed from self-model prediction error, weighted by self-relevance, and modulated by prior affective states and by what we term temporal aiming, which captures affective appraisal over time. This accommodates shifts in the self-other boundary, allowing affect to emerge at individual and dyadic levels. We propose a novel method termed geometric hyperscanning, based on the Forman-Ricci curvature, to operationalize these processes empirically: it tracks topological reconfigurations in inter-brain networks, with its entropy serving as a proxy for affective phase transitions such as rupture, co-regulation, and re-attunement.

\keywords{Active inference \and Hyperscanning \and Second-person neuroscience \and Dyadic coupling \and  Forman-Ricci curvature \and Phase transitions.}
\end{abstract}

\section{Introduction}

How do agents come to feel ‘together’? How do they navigate the sharing of beliefs about the world and themselves, and the fragile terrain of affective meaning? Traditional models of social cognition often approach this question through the lens of mental state attribution: one agent models another’s beliefs, intentions, and emotions. Adopting a third-person stance towards someone has dominated theoretical and empirical social cognition research \cite{catal2024belief}. However, such models ignore a crucial fact: social understanding, in most of its ecologically valid forms, does not unfold in detached observation but in active, embodied engagement with others. As Schilbach et al. \cite{schilbach2013secondperson} argue, social interaction is not simply the backdrop against which inference occurs; instead, it is the medium through which meaning is constituted. This shift, often captured under the rubric of `second-person neuroscience' \cite{redcay2019secondperson,schilbach2025synchrony}, reframes the unit of analysis from the isolated agent to the dyad as a generative system in its own right, by taking mutual engagement in social interaction, rather than social observation, as its central explanandum. Recent formalizations of this framework through the lens of active inference \cite{lehmann2023secondperson} have provided the missing computational scaffolding: they model interacting agents as mutually coupled generative systems that co-regulate meaning via recursive belief updating. In this view, social understanding is not computed about another agent, but emerges while in social interaction with another through ongoing cycles of expectation, violation, and realignment. The dyad becomes a unit of analysis in its own right and can be regarded as a shared generative manifold, which can be modeled as a coupled system of intra- and interpersonal processes across multiple modalities \cite{bolis2017dialectical,bolis2023attunement}.

\subsection{The Affective Gap}

We extend second-person active inference to the domain of affect, modeling it as recursive inference over two types of prediction error: mismatch between predictions and outcomes relevant to the world, and those relevant to identity. Drawing on Jiang and Luo's \cite{jiang2025valence,jiang2024jointembedding} model, valence reflects an inference over the integrity of the self-model, embedded in the dyadic generative manifold. Because both agents maintain self-models while recursively modeling each other, affective dynamics become entangled, shaping belief updating and action selection. This recursive process promotes interpersonal attunement and stabilization \cite{bolis2023attunement,vanvugt2020synchronization,scholtes2020synchrony}, while affective rupture triggers cognitively costly recalibration \cite{christovmoore2025chills,albarracin2024selfesteem}. We propose geometric hyperscanning to link these formal dynamics to neural signatures, expanding the framework of second-person active inference and its formal dynamics of belief and affect to empirical research.\\

This paper makes three contributions. First, we formalize affect as recursive inference over self-model coherence. We model valence as identity-relevant prediction error weighted by self-relevance, modulated by affective memory, and shaped by temporal aiming --- the agent's orientation across past and future affective states (Section 2.1). We show how recursive modeling of the other dynamically shifts self--other boundaries (Section 2.2). Second, we introduce geometric hyperscanning, an empirical method based on Forman-Ricci curvature (FRc), to track topological reconfigurations in inter-brain networks and infer affective phase transitions (Section 2.3). Third, we integrate this formal-empirical framework within second-person neuroscience and active inference. We offer a scalable architecture for modeling recursive affective dynamics across psychotherapy, development, and naturalistic interaction, and outline its broader implications (Section 3).

\section{Second-Person Neuroscience and Dyadic Active Inference}

Second-person neuroscience \cite{schilbach2013secondperson} holds that social understanding is constituted through real-time mutual engagement, as opposed to detached or post-hoc attribution. From a computational standpoint, this view is well-expressed in the active inference framework, in which agents minimize variational free energy to maintain a coherent embodied model of the world and their role within it. In second-person settings, this framework must be extended \cite{bolis2020interact,lehmann2023secondperson,veissiere2020minds}: agents not only predict external states, but also the inferred generative models of others. These recursive beliefs form a shared generative manifold, wherein action and perception are jointly coordinated. We can conceive of each agent maintaining a generative model:
\begin{equation}
M = p(o_i, a_j, s_i, s_j \,|\, \pi_i)
\end{equation}
where:
\begin{itemize}
\item $o_i$ = observations available to agent $i$,
\item $a_j$ =  inferred actions of agent $j$,
\item $s_i, s_j$ = hidden states of agents $i$ and $j$,
\item $\pi_i$ = current policy or sequence of expected actions for agent $i$.
\end{itemize}

The recursive nature of this system (wherein both agents model each other, modeling them) does not lead to an infinite regress. Instead, as Friston and Frith (2015)  demonstrated, it dissolves into a shared dynamic narrative, where both agents come to predict and enact the same process. In other words, both agents use the same model and the dyad, thus, becomes the minimal unit of inference.

In contrast, traditional third-person approaches to social cognition, such as Theory of Mind or mindreading paradigms, cast social understanding as an observer inferring another’s hidden mental states. Classic experiments, for instance, might have a participant watch a video or read a story and then guess what another person believes or intends. These paradigms, while valid, involve a one-way attribution process, often in non-interactive settings. Neuroscientifically, third-person “mentalizing” tasks reliably engage a network including the dorsomedial prefrontal cortex (dmPFC), ventromedial prefrontal cortex (vmPFC), and temporoparietal junction (TPJ), which is often called the mentalizing or theory-of-mind network \cite{gvirts2020commentary}. However, in a live interaction, one’s brain is modeling the other and responding to them in real time, creating a feedback loop. This form of social connectedness has measurable two-brain dynamics: an engaged “I–Thou” exchange produces alignment of neural rhythms that does not occur in a one-directional “I observe him” situation (see \cite{dumas2010inter,schilbach2025synchrony} for a recent review).

\cite{gvirts2019neurobiological} refers to these coupled networks as mutual social attention systems, proposing that when two people directly attend to each other, their brains temporarily form an integrated system. The second-person emphasis on real-time mutual engagement dovetails with active inference, because both paradigms recognize that understanding others is an active, dynamical, and bidirectional alignment process.

\subsection{Valence as Inference over the Self-Model}

Within this coupled generative architecture, affect is conceptualized as dynamic regulatory processes guiding cognition and behavior based on the integrity and coherence of an agent’s predictive self-model \cite{jiang2025valence,jiang2024jointembedding}. Central to this approach is the idea that emotional valence signals the alignment or mismatch between predicted and observed outcomes, particularly those relevant to identity and social expectations \cite{albarracin2024selfesteem}.

Prior work has modeled valence as a derivative of free energy \cite{joffily2013valence}, a metacognitive signal \cite{hesp2021valence}, interoceptive inference \cite{smith2019emotions}, shifts in processing modes \cite{yanagisawa2023valence}, and recursive affective dynamics linked to self-relevance and identity \cite{albarracin2023husserl,jiang2025valence,albarracin2024selfesteem}. Building on these insights, and drawing on \cite{jiang2025valence}, we propose an extended model of emotional valence as a continuous inference over the integrity of the agent’s self-model. Emotional states emerge from prediction errors tied to identity-relevant expectations. Valence operationalizes the emotional appraisal of interactions via two primary factors: (1) the magnitude of self-model prediction error, and (2) the self-relevance assigned to that error. We further incorporate a temporal parameter into this model, reflecting the directionality and velocity of affective processing. Building on \cite{joffily2013valence} and \cite{yanagisawa2023valence}, we introduce \textit{temporal aiming} (see Fig. 1), encompassing both the velocity of affective evaluation (fast learning rates vs. slow enduring states) and its temporal direction (retrospective vs. prospective appraisal).

\begin{figure}[ht]
\centering
\includegraphics[width=0.8\textwidth]{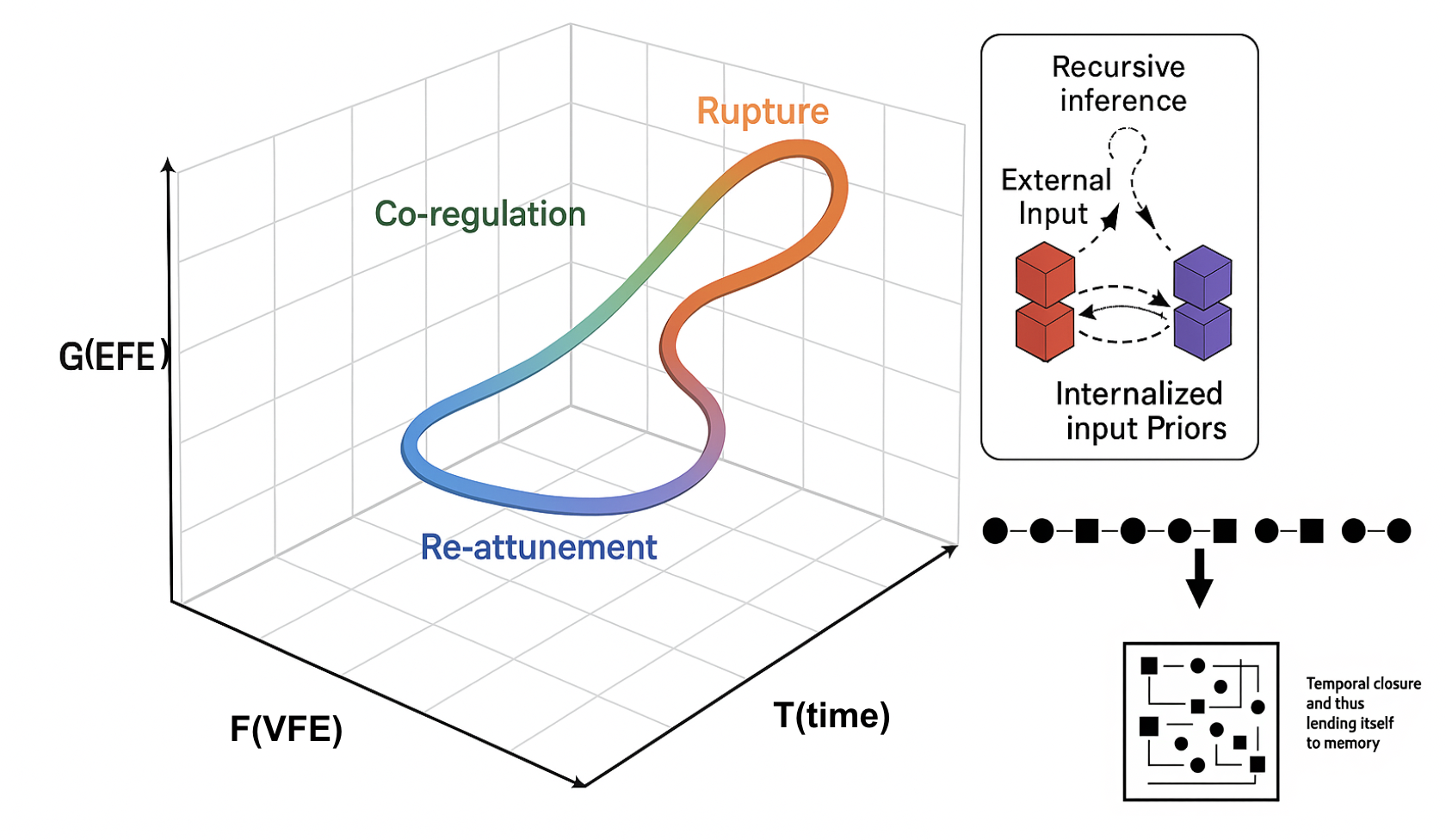}
\caption{\textit{Temporal aiming in affective inference.} The figure illustrates how agents evaluate self-model prediction error over time according to different temporal orientations. Affect is shaped not only by the magnitude and relevance of prediction error, but also by the agent’s temporal aim, i.e., whether inference is directed toward anticipated futures or prior experiences, and whether updates occur rapidly or gradually. This temporal structuring of affect modulates the stability and flexibility of policy selection within dyadic interaction.}
\end{figure}

This modulation adds an essential dimension to affective inference: agents may respond differently to the same interaction depending on whether they are prospectively anxious, retrospectively regretful, or temporally stable. Temporal aiming thus complements self-relevance and prediction error as core variables in our valence model. We conceptualize emotions as functions of valence (alignment of identity-relevant predictions) and arousal (uncertainty and intensity of self-relevant evaluations), with both dimensions anchored in the evolving self-concept, shaped by reflection, comparison, and internalized ideals. In this framework, valence reflects the agent’s appraisal of ongoing interaction or how “good” or “bad” the current moment feels to the agent, computed as a weighted function of self-model prediction error and self-relevance, and modulated by prior affective memory \cite{jiang2024jointembedding}:
\begin{equation}
V_t = \alpha \cdot PE_{\text{self},t} \cdot SR_t + \beta \cdot V_{SR,t}
\end{equation}

where:
\begin{itemize}
  \item $PE_{\text{self},t}$ is the prediction error between expected and actual self-state, capturing the “surprise” the agent experiences about itself
  \item $SR_t$ is a dynamic scalar indexing the self-relevance of the context, reflecting how much the situation matters for the agent’s identity
  \item $V_{SR,t}$ is retrieved from similar prior contexts, representing the lingering emotional tone (mood) shaped by related past experiences,
  \item $\alpha, \beta$ are precision-modulating weights, that determine how strongly the agent reacts to immediate surprises versus how much past mood carries over..
\end{itemize}

It follows that whenever the agent experiences a self-related prediction error, it generates an affective response proportional to that error and to how self-relevant it is. The first term, $\alpha \cdot PE_{\text{self},t} \cdot SR_t$, encodes the instantaneous emotional impact: large, self-relevant surprises cause large valence shifts. The parameter $\alpha$ scales the impact of immediate prediction error on valence. An agent $i$ with a larger $\alpha$ will exhibit heightened emotional sensitivity to surprises, particularly negative ones, which yield more negative valence. The second term, $\beta \cdot V_{SR,t}$ with $0 < \beta < 1$, he second term, $\beta \cdot V_{SR,t}$, introduces emotional inertia: it blends in past affective states, so that current mood is partly a continuation of recent emotional history, it makes valence a \emph{leaky integrator}: it carries forward some fraction of previous valence (mood) into the current moment. The term $PE_{\text{self},t}$ includes both the agent's previous valence $V(t - 1)$ and contributions from past episodes that had similar self-relevance or context as the current one.  We can treat $V_{SR,t}$ as an exponentially decaying sum of past valences, with greater weight assigned to more recent times and to events with comparable self-relevance (SR). In the notation of the original model, self-model prediction error and self-relevance map onto $PE_{\text{self}}$ and $SR$, respectively, where the last term of $V_{SR,t}$ helps to maintain the affective experience as iterative and continuous.

\subsubsection{Valence under the Dyadic Inference Framework}

In dyadic active inference, each agent’s self-model is treated as a dynamic function of its internal world model, which includes a simulated model of the other agent and incoming interactional feedback. At time $t$, agent $i$’s self-model is conditioned on beliefs about the other agent’s state, actions, and observations:
\begin{equation}
P^i_t = f(S^i_t, A^j_t, O^j_t)
\end{equation}

This reflects that one’s self-concept in a social setting is co-constructed by self-generated and socially observed information.
Agent $i$’s full generative model factorizes as:
\begin{equation}
P(O_i, A_i, S_i, S_j) = P(O_i\,|\,S_i, A_i)\cdot P(A_i\,|\,S_i, S_j)\cdot P(S_i\,|\,S_j)
\end{equation}

This expresses that (1) the agent’s observations depend on its state and actions, (2) its actions depend on both its own and the other’s state, and (3) its own state is partly inferred from the other’s state. Valence ($V_i, t$) measures how emotionally coherent the agent’s self-model feels over time. It is computed from the prediction error in the self-model, modulated by both $SR_t$ and $V_{SR,t}$. The expected valence is defined as:

\begin{equation}
\mathbb{E}(V_i^t) = \alpha \cdot PE_{\text{self},t} \cdot SR_t + \beta \cdot V_{SR,t}
\end{equation}

 where $PE^i_{\text{self},t} $ is the prediction error between expected and actual self-state, and
alpha and beta are precision-modulating weights. The realized valence reflects how well
observed outcomes under the current policy$ pi_{i}^{t}$ align with these expectations; it is
given by:
\begin{center}    
\begin{align}
V_i^t &= \mathbb{E}_{P(o^t \mid \pi^t)} \left[ \log P\left(o^t \mid \mathbb{E}(V_i^t) \right) \right] \notag \\
      &= \mathbb{E}_{P(o^t \mid \pi^t)} \left[ \log P\left(o^t \mid \alpha \left|  P_i^t - \hat{P}_i^t \right| SR_i^t + \beta V_{i,SR}^t \right) \right]
\end{align}
\end{center}

This means that valence is the expectation, over possible outcomes, of how well those outcomes will sustain the agent’s internalized social narrative, given its evolving model of the partner and the feedback from ongoing interaction.

% the environments 'definition', 'lemma', 'proposition', 'corollary',
% 'remark', and 'example' are defined in the LLNCS documentclass as well.
%

\subsubsection{Affective Valence as a Prior Over Policy in Planning-as-Inference}

By design, valence serves as an internal reward/safety signal: high positive valence indicates that the interaction is proceeding smoothly, with strong affective synchrony, while negative valence signals a breakdown in synchrony, prompting either a new attempt to restore alignment or a withdrawal from the interaction, depending on individual differences in valence construction and perception (modeled by the parameters $\alpha$ and $\beta$).
Which path the agent takes depends on individual differences in emotional sensitivity and persistence, captured by the parameters $\alpha$ and $\beta$
Though we do not explicitly model all downstream effects here, valence functions as a feedback loop, a readout of the self-model’s performance and modulating subsequent perception and action. In practical terms, this means that affect shapes the space of possible future actions by making certain policies feel “more natural” or “safer” than others. In our model, affect modulates the posterior over policies, effectively serving as a soft prior that biases action selection toward more affectively coherent trajectories. Following Da Costa's (2020) \cite{dacosta2020discrete} planning-as-inference paradigm, we express policy selection as:
\begin{equation}
P(\pi_i\,|\,V_t) \propto e^{\lambda \cdot g(V_t, \pi_t)}
\end{equation}
where:
\begin{itemize}
\item $\lambda$ is an inverse temperature parameter controlling affective precision or confidence in the valence signal (high 
$\lambda$ = more deterministic choice; low 
$\lambda$ = more exploratory choice), and
\item $g(V_t, \pi_t)$ is a compatibility function mapping affective coherence to policy likelihood (higher compatibility = higher policy probability).
\end{itemize}

This captures the recursive and value-sensitive nature of planning in social contexts. The agent is more likely to choose plans consistent with positive valence (i.e., minimal surprise and maximal narrative alignment), while negative valence reduces the likelihood of maintaining the current trajectory. When both agents are engaged in this recursive affective inference, the system forms a \textit{dynamically coupled manifold} —a shared generative space in which each agent’s self-coherence is entangled with the other's behavior. Moments of affective rupture mark local minima in joint narrative alignment and correspond to dyadic expected free energy peaks. Repairing these ruptures requires coordinated adjustments in beliefs and policies across both agents. This framework completes a recursive loop: expected valence, derived from self-model integrity and anticipated partner behavior, acts as a prior over policy selection, guiding the agent toward actions expected to preserve or restore affective coherence. These chosen policies lead to new interactions and observations, which update the agent’s self-model and generate realized valence. In turn, the mismatch between expected and realized valence yields an affective prediction error that shapes the next inference cycle. This recursive loop (Fig. 2) summarizes how affectively modulated policies guide dyadic interaction through cycles of prediction error, self-model update, and realignment.
\begin{figure}
\centering
\includegraphics[width=0.9\textwidth]{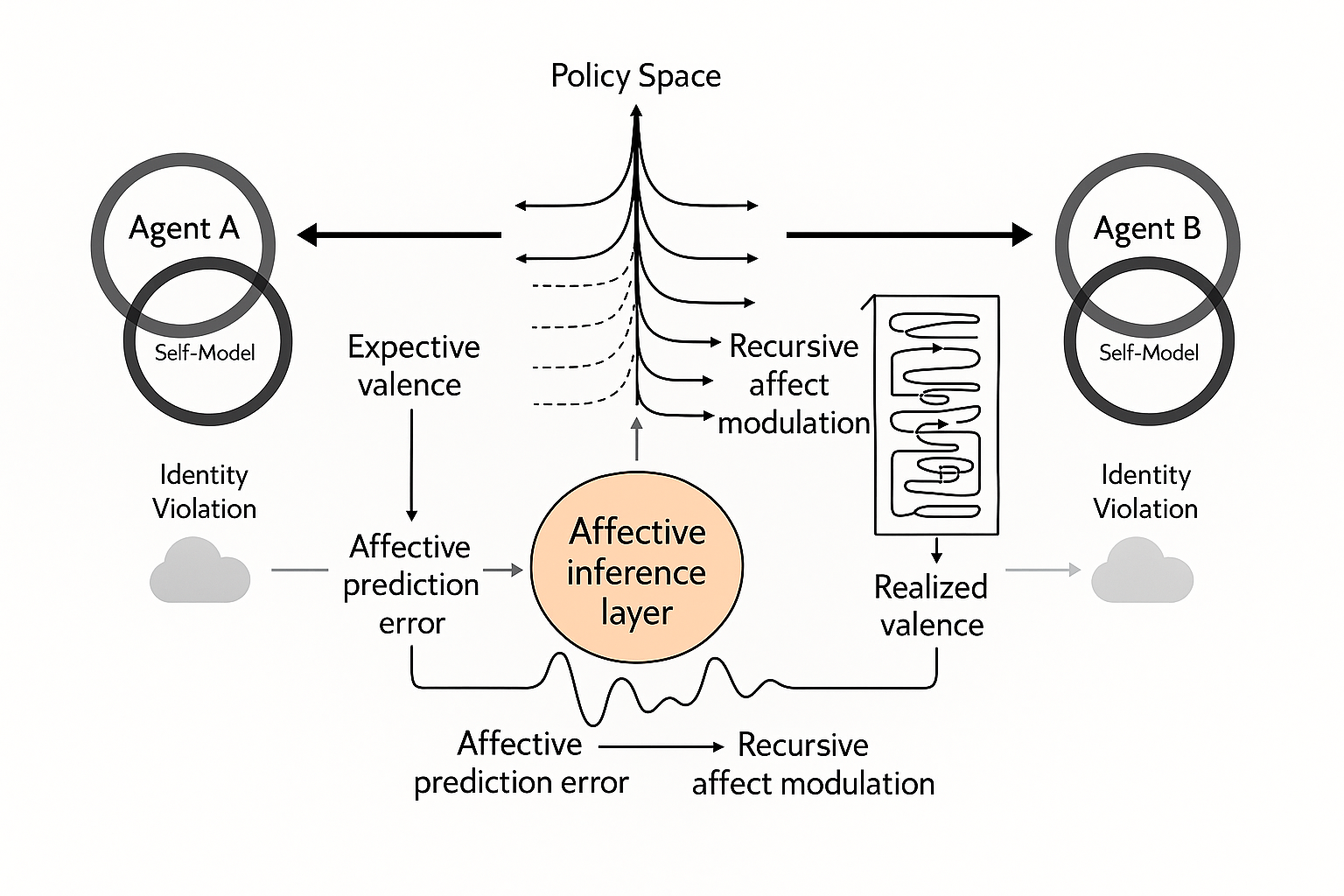}
\caption{ \textit{Recursive Affective Inference Framework for Dyadic Social Interaction.} Illustrates a recursive framework wherein affect dynamically regulates social behavior between two agents via cycles of inference, prediction error, and adaptive policy updates grounded in self-model coherence.}
\label{fig:recursive_framework}
\end{figure}

\subsection{Recursive Coupling and Shared Surprise}

In exploring the dynamics of affective interactions, it is helpful to draw upon the concept of intra-action from neo-materialist philosophy \cite{barad2007meeting}. Intra-action emphasizes that interacting entities do not preexist independently but emerge through mutual entanglement. Under this lens, the dyadic interaction is not simply a transfer of emotional states between two separate agents; instead, it represents a continual co-construction of affective experiences and identities \cite{rahmjoo2023intraactive}. From this perspective, affective states are inherently recursive and relational, arising from the iterative feedback loops between agents. Affective ruptures, moments of misalignment or increased prediction error, serve as significant informational signals indicating points of divergence in the shared expectations of interacting agents. These affective evaluations modulate the posterior over policies, biasing action selection toward trajectories that restore narrative alignment and reduce identity-relevant surprise. Such an entanglement of affect and policy selection turns the dyad into a dynamic field of co-regulated affective inference, where both agents are simultaneously minimizing joint expected free energy:

\begin{equation}
G = \mathbb{E}_{q(s)}[\log q(s) - \log p(o, s)]
\end{equation}

Then, in a dyadic setting, we can express the joint free energy as:
\begin{equation}
G_{\text{joint}} = G_i + G_j = \mathbb{E}_{q(s_i, s_j)}\left[\log \frac{q(s_i, s_j)}{p(o_i, o_j, s_i, s_j)}\right]
\end{equation}

Here, 
$G_{\text{joint}}$ quantifies the shared divergence between the dyad’s joint beliefs and its joint generative model. High joint valence corresponds to low joint free energy (smooth alignment), whereas affective ruptures correspond to spikes in joint free energy (misalignment and uncertainty).
At the dyadic level, recursive coupling of valence creates a shared attractor structure: higher valence biases policy selection toward trajectories that preserve narrative coherence and reduce expected future free energy. Conversely, negative valence can trigger exploratory policies to restore alignment or withdraw from destabilizing interactional trajectories. This way, valence modulates immediate affective dynamics and the agent’s future epistemic and instrumental uncertainty.
\paragraph{}
Critically, moments of affective rupture—sharp drops in valence—correspond to increases in joint free energy, typically arising from mismatches in the agents’ self-model priors. These transitions are not noise but informative inflection points in the dyadic generative process. Mismatches in self-model coherence directly elevate free energy, as valence encodes deviations between observed interaction outcomes and expected self-relevant predictions. Because valence reflects each agent’s affective confidence in their internal model, it serves as both a readout of local free energy minimization and a control signal that shapes future joint free energy via policy adaptation. This recursive structure is central to how dyadic affective inference maintains, destabilizes, or reconstructs interpersonal synchrony over time.

\subsection{Geometric Hyperscanning as Second-Person Method}

To empirically access the internal dynamics of dyadic affective inference, we introduce \textit{geometric hyperscanning} based on the Forman-Ricci curvature (FRc) \cite{hinrichs2025markers,hinrichs2025geometryinterbrainnetworks}, which captures phase transitions in inter-brain network topology using geometric markers derived from EEG hyperscanning data. This method enables operationalization of latent affective and narrative coherence as measurable transformations in the topology of inter-brain networks. Importantly, FRc is not treated as a post hoc readout; we propose it as a data-derived proxy for latent generative states in the dyadic active inference model. Traditional hyperscanning methods focus on node-level synchrony, typically correlating homologous regions across two brains. While informative about temporal alignment, these methods fail to detect topological reconfigurations in inter-brain structure. FRc, by contrast, is an edge-centric geometric measure that quantifies the local expansion or contraction of topological flow within a network \cite{weber2017curvature}, and has been shown to characterize network robustness, signal routing, and functional reorganization.
\paragraph{}
Given a network $G = (V, E)$ with edge $e_{ij}$ connecting nodes $v_i$ and $v_j$, the FRc is computed as:
\begin{equation}
\text{FRc}(e_{ij}) = w(e_{ij})\left(\frac{1}{w(v_i)} + \frac{1}{w(v_j)}\right) - \sum_{e \sim e_{ij}} \left(\frac{w(e_{ij})}{\sqrt{w(e_{ij}) \cdot w(e)}}\right)
\end{equation}
where we interpret $w(\cdot)$ as node and edge weights (e.g., derived from functional connectivity matrices), to track the evolution of inter-brain topology over time, we calculate the entropy of the FRc distribution within a sliding window.

\begin{equation}
H_{\text{FRc}}(t) = - \sum_{e \in E_t} p(e) \log p(e)
\end{equation}

This scalar quantity acts as a proxy for interactional volatility. Peaks or discontinuities in FRc entropy correspond to phase transitions in the shared generative manifold and reflect affective rupture, repair, or co-regulation dynamics.
\begin{equation}
p(e) = \frac{\text{FRc}(e)}{\sum_{e'} \text{FRc}(e')}
\end{equation}

FRc entropy is not simply an output variable but can be formalized as an observation model over latent affective prediction error. Curvature dynamics can infer or constrain internal state variables such as valence or identity coherence. This closes the loop between model and measurement, as the generative model predicts changes in self-model coherence and dyadic free energy, and the FRc entropy serves as an empirical signal that can be used to update model beliefs via Bayesian inference. To make these transitions tractable and interpretable, we propose a multi-level operationalization of rupture, repair, and reattunement, linking neural signatures to behavioral and psychological indicators. Neurally, rupture corresponds to abrupt drops in network integration (low FRc), high curvature entropy, or rapid shifts in entropy gradients; repair and reattunement reflect a return to more stable topologies. Behaviorally, rupture may manifest as disengagement, disrupted synchrony, or confusion (e.g., gaze aversion, vocal interruptions), while repair involves re-engagement cues such as gesture mirroring, prosodic modulation, or turn-taking; reattunement is evidenced by fluent, coordinated interaction and mutual responsiveness. Psychologically, rupture can be self-reported as disconnection or emotional distancing, repair as subjective efforts to reconnect, and reattunement as restored mutual understanding or relational safety. These experiences can be measured via self-reports (affective sliders, post-session interviews, validated instruments), annotated through multimodal behavioral data (gaze, prosody, synchrony), and aligned with curvature dynamics in the inter-brain manifold. Treated as latent categorical states (e.g., within a hidden Markov model), these phases can serve as points of annotation, classification, or real-time feedback in both experimental and clinical contexts.
\paragraph{}
Importantly, embedding these events within the recursive generative model of dyadic inference allows rupture–repair cycles to function as outcomes and as endogenous features of interactional dynamics. Affective coherence drives policy selection, while behavioral and neural feedback recursively update each agent’s self-model. This recursive structure enables in situ inference over dynamic interpersonal states, validation of model predictions across modalities, and fine-grained adaptation to moment-by-moment shifts in relational coherence. Integrating \textit{geometric hyperscanning} into the generative model therefore bounds the model with an additional objective: not only minimising identity-relevant prediction error, but implicitly driving the system toward maximal interpersonal attunement—an intrinsically rewarding state even in the absence of overt instrumental gain \cite{redcay2019secondperson,bolis2020interact}. Curvature-induced surprises can surface to awareness, prompting explicit attempts at repair.
\paragraph{}
In sum, FRc-based \textit{geometric hyperscanning} offers a scalable, multimodal method for detecting and interpreting recursive affective inference. By triangulating neural entropy, behavioral dynamics, and psychological experience, we gain an integrated empirical handle on how dyads co-regulate meaning, affect, and identity over time. This framework advances a formal, testable architecture for second-person neuroscience and lays the groundwork for modeling recursive emotional entanglement, the subject of the following section.

\section{Future Directions and Conclusions}

This paper proposed that affect is best understood as recursive inference over self-model coherence within coupled generative systems. Modeling the dyad via active inference, we formalized affect as a valuation of identity-relevant surprise, driving and modulating recursive loops of belief updating, policy selection, and behavioral adaptation during social interaction. We introduced a novel method based on the FRc, termed \textit{geometric hyperscanning} \cite{hinrichs2025geometryinterbrainnetworks}, which tracks dynamic reconfigurations in inter-brain networks, providing an empirical window onto moments of rupture, repair, and re-attunement. By unifying formal models of belief dynamics, affective evaluation, and network geometry, we advance a scalable and interpretable framework for operationalizing second-person active inference.

\subsection{Future Directions}

The proposed theoretical formulation invites several concrete avenues for empirical investigation and computational development. Foremost among these is the simulation of agent-based models in which curvature entropy is treated as a streamed sensory input—continuously informing belief updates about narrative alignment. Such models would allow us to test whether the architecture, under active inference, spontaneously reproduces the rupture–repair cycles observed in real dyadic interaction. In doing so, they provide a crucial bridge between formalism, neurophysiology, and phenomenology. 
Beyond dyads, the framework naturally extends to hierarchically structured systems. By introducing group-level alignment variables, one can model small-group coherence as an emergent property of multiple interacting pairwise dynamics. This multiscale extension opens the door to collective behaviour, organisational synchrony, and crowd psychology applications, where relational inference is distributed across agents and temporal scales. Furthermore, the curvature-informed likelihood may be productively integrated with other sensorimotor channels. Prosodic variation, gaze dynamics, and even facial microexpressions can all be incorporated within a deep generative model, allowing heterogeneous evidence streams to converge on shared latent variables. This multimodal fusion enriches the inferential landscape and supports more robust decoding of the intersubjective field.
Yet the prospect of real-time inference over relational states is not without ethical consequences. As structural priors become actionable in applied contexts—be it psychotherapy, education, or social robotics—the issues of privacy, autonomy, and interpretability move to the fore. Any translational implementation must therefore be grounded in transparent consent procedures and designed to yield intelligible feedback to clinicians and researchers and to the individuals whose relational trajectories it purports to track.

\subsection{Sociomarkers for Interpersonalized Psychiatry}

\textit{Geometric hyperscanning} offers a promising foundation for \emph{sociomarkers}: real-time, measurable signatures of dynamic interpersonal coordination \cite{bolis2023attunement}, analogous to biomarkers of individual physiology. Affective co-regulation---vital for social bonding across species \cite{coleman2021duet,fortune2011duet,brosnan2003inequity,roma2006monkeys}---is foundational to human development, shaping cognition, emotion, and relational selfhood through early caregiver-infant interactions \cite{bolis2020interact,fini2023roots,vygotsky1978mind,fotopoulou2017homeostasis,hoehl2018secondperson}. Misattunement and repair cycles scaffold resilience and relational growth \cite{bolis2023attunement,bolis2020interact}, with fluctuations in curvature entropy potentially indexing key developmental processes \cite{schilbach2013secondperson,cittern2018attachment,wass2020entrainment}. Such signatures could inform personalized \cite{mandelli2023autism,lombardo2019bigdata} and interpersonalized psychiatry \cite{bolis2023attunement}, which views psychopathology as a disruption in co-regulation and narrative repair \cite{bolis2017dialectical,milton2012doubleempathy}. In psychotherapy, real-time dyadic metrics \cite{bolis2017dialectical,lahnakoski2020tracking} can track attunement, rupture, and repair \cite{ramseyer2011synchrony}, helping both therapists and clients navigate the relational dynamics of change. Crucially, synchrony is neither inherently good nor constant: transitions into and out of synchrony may reflect autonomy, resistance, or meaningful differentiation, not dysfunction. Modeling these dynamics as recursive affective inference provides a principled lens for basic science and relational mental health care.

\subsection{Conclusions}

Building on these formal and empirical insights, this approach carries several key implications. First, it invites a rethinking of affect in social interaction—not merely as a byproduct of inference, but as a primary regulatory signal that modulates the stability and flexibility of generative coupling itself. On this view, affect becomes the sense of coherence or incoherence within the dyadic narrative, indicating when beliefs must be renegotiated or when interactional roles require repair.
Second, by shifting emphasis from synchrony per se to the topology of inter-brain networks, our method aligns with a broader movement in second-person neuroscience toward characterizing interaction structure. In this respect, FRc extends Friston and Frith’s (2015) \cite{friston2015duet} notion of generalized synchrony by capturing the reconfigurations in network structure that accompany recursive affective inference.
Third, by embedding affect into the posterior over policies, we extend da Costa’s (2020) planning-as-inference formalism into the affective domain. Crucially, the affective signal can operate implicitly, as a precision-weight on policy beliefs that never reaches phenomenal awareness, or explicitly, as a consciously accessible feeling that furnishes the agent with propositional knowledge about its own relational stance. In either guise, policy selection is driven not only by epistemic or instrumental imperatives, but also by the imperative to preserve narrative and relational coherence. Affect thereby becomes an intrinsic dimension of social planning, covertly biasing action when tacit and overtly guiding behaviour when made conscious—steering agents through moments of ambiguity toward restored interactional alignment.
These considerations open a range of empirical and computational possibilities. In psychotherapy, curvature entropy may serve as a real-time sociomarker of rupture and repair; in early development, FRc could illuminate how co-regulatory patterns emerge between infants and caregivers. Simulations of affective agents equipped with recursive self- and other-models offer a formal means to investigate relational resilience and breakdown. In particular, such models allow us to test the dialectical misattunement hypothesis (Bolis et al., 2017), which posits that psychiatric risk arises not from individual deficits alone but from persistent mismatches in communicative expectations or learning rates. Simulating these mismatches under controlled conditions reveals how dyads adapt—or fail to—over developmental time, providing insight into vulnerability and repair mechanisms.
The theoretical landscape also invites expansion. How does this framework scale to group interactions, where recursive generative coupling involves multiple agents? What ethical safeguards are needed when automating the detection of affective rupture and repair—especially in sensitive contexts like therapy or education? And how might shared belief geometry interact with linguistic, prosodic, and bodily cues to form a rich, multimodal map of affective inference?
Together, these questions point toward a broader scientific project: one that integrates formal modeling, empirical measurement, and philosophical reflection into a unified account of relational meaning. In this view, the dynamics of belief, affect, and participation are not peripheral but fundamental to the fabric of social cognition. By reframing affect as recursive inference within coupled generative systems—and rendering those dynamics tractable via network geometry—we move closer to the vision of interaction as the basic unit of social neuroscience. Our proposal thus offers both a conceptual framework and methodological toolkit for studying human interaction's structural and affective choreography.

\begin{credits}
\subsubsection{Contributions.} 
NH and YJ jointly conceived and drafted the initial manuscript. MA and DB contributed core conceptual ideas and critically revised the text. LCM and LS provided additional feedback and editorial refinement. All authors reviewed and approved the final version of the manuscript.

\subsubsection{\ackname} 
NH and YJ would like to thank Danielle J. Williams and the inaugural meeting of the Society for Neuroscience and Philosophy, at which the authors contributed papers and discussed integrating their presented approaches. NH would like to thank Lancelot Da Costa and Sebastian Sosa for their insightful feedback.

\subsubsection{\discintname}
 NH was funded by CNPq, MPI-CBS, and OIST. DB receives funding from the IIT. LS receives funding from the DFG.
\end{credits}
\\
\appendix

\section{A Simulation-Ready Implementation (MBSR $\,+\,$ Valence–HMM)}
\label{app:hmm-proposal}

\paragraph{Goal.}
We instantiate the manuscript’s account with two coupled but distinct components: (i) an autobiographical memory-based self-relevance, (ii) a \emph{valence} HMM that carries the temporal dependence of affect. The HMM framework is motivated by the temporal persistence of valence exhibiting state continuity with gradual, experience-driven change (for review: affective spillover/carryover effect). HMM transition matrices capture this through high diagonal probabilities while permitting off-diagonal transitions. Standard moment-by-moment computations lack this persistence; HMMs provide it via probabilistic state evolution.

\subsection{Setup and notation}
For agent $i$ and step $t$, the internal model $M_i$ maintains a self–state prior/posterior $P_i^{t},\hat P_i^{t}$, observations $o_i^{t}$, inferred partner actions $a_j^{t}$, and the identity prediction error
\[
PE_{\text{self},i}^{t}=\hat P_i^{t}-P_i^{t}.
\]

The projected/predicted valence, by definition, uses a scalar self-relevance $SR_{t}^i\in[0,1]$ and the identity prediction error $PE_{\text{self},i}^{t}$ to weight identity-relevant surprise:

\begin{equation}
\label{eq:valence}
\mathbb{E}(V_i^t) = \alpha \cdot PE_{\text{self},t} \cdot SR_{t} + \beta \cdot V_{SR,t}
\end{equation}
Here $\alpha,\beta$ are precision-modulating gains balancing instantaneous identity-relevant surprise and affective carryover $V_{t, SR}^i$.
\\
The agent’s current policy $\pi_i^t$ is enacted based on the valence state:
\begin{equation}
\label{eq:policy}
P(\pi_i\,|\,V_t) \propto e^{\lambda \cdot g(V_t, \pi_t)}
\end{equation}
where $\phi(\cdot)$ may include linear terms and selected interactions; its purpose is to summarize, per trial, the stimulus and self-context the agent brings to the exchange.

After one carries out an action policy, the results from the action will be made available as new observations to compute the realized valence $V_i^t$. By comparing expected outcomes under the current policy to the expectation $\mathbb{E}(V_i^t)$, this item formalizes the felt value of the present exchange as the log-likelihood of observations given the agent’s valence-weighted model (Eq.~\ref{eq:valence-realized-app}).  

\begin{align}
\label{eq:valence-realized-app}
V_i^t
&=
\mathbb{E}_{P(o^t \mid \pi_i^t)}
\!\left[
\log P\!\left(
o^t \,\middle|\, \mathbb{E}(V_i^t)
\right)
\right].
\end{align}

In the following implementation, we discretize both self-relevance and valence into three ordered HMM states:
\begin{itemize}
  \item \text{Self–relevance (SR):} $R_i^t \in \{\,r_{\mathrm{low}},\, r_{\mathrm{med}},\, r_{\mathrm{high}}\,\}$.
  \item \text{Valence (S):} $S_i^t \in \{\,s_{\mathrm{neg}},\, s_{\mathrm{neu}},\, s_{\mathrm{pos}}\,\}$.
\end{itemize}

\subsection{Memory-based Self–Relevance Chain}

For each interaction, agents encode the present exchange in an interaction descriptor
\begin{equation}
z_i^{t}=\big(o_i^{t},\,a_j^{t},\,P_i^{t}\big)\in\mathbb{R}^d
\label{eq:z}
\end{equation}
\label{app:mem-sr}
Each agent maintains an autobiographical store $\mathcal{LTM}_i=\{(x_k,I_k,\tau_k)\}_{k=1}^K$ where each memory $m_k$ has a descriptor $x_k$ (context features), an
impact tag $I_k \in [0,1]$ (how strongly the episode affected the agent), 
and a timestamp $\tau_k$.
Given the current context $z_i^t$, we compute how similar it is to each
stored memory. These similarities are then normalized so that they add up to one, turning
them into attention weights over memory:

\[
Sim_k^t=\mathrm{sim}(z_i^t,x_k),\qquad
\tilde Sim_k^t=\frac{Sim_k^t}{\sum_j Sim_j^t},
\]
For each memory  $m_k$ stored, the context $z_i^t$ will become the $x_k$, the time becomes $\tau_k$, and they are accompanied by an impact tag which represents (1) how large the self-prediction error was, and (2) how much the policy
distribution reorganized. Both are normalized to $[0,1]$ for comparability. Impact tags are written from the realized change at step $t$ using a bounded, divergence-free score:
\begin{align}
I^t &= \mathrm{norm}\!\big(\|PE_{\text{self},i}^{\,t}\|\big)\;\times\;\mathrm{TV}_{\pi}^{\,t}, 
\label{eq:impact}\\
\mathrm{TV}_{\pi}^{\,t} &= \tfrac12 \sum_{\pi\in\Pi_i}\big|p_{\text{post}}^{\,t}(\pi)-p_{\text{prior}}^{\,t}(\pi)\big|\in[0,1],
\label{eq:tv}
\end{align}
where $p_{\text{prior}}^{\,t}$/$p_{\text{post}}^{\,t}$ are policy distributions (softmax over scores) before/after assimilating step $t$; $\mathrm{norm}(\cdot)$ maps magnitudes to $[0,1]$.

A \emph{memory-derived relevance score} is then obtained as a similarity-weighted average of impact tags:
\begin{equation}
\widetilde{SR}_t^i=\sum_{k=1}^K \tilde Sim_k^t\,I_k \in[0,1].
\label{eq:SR-memory}
\end{equation}

\subsection{Expected Valence HMM Chain}
\paragraph{Expected valence.}
We first compute the expected valence as a product of the previous step's
identity prediction error and self-relevance (Eq.~\ref{eq:valence}).  

\paragraph{Integration into the valence HMM.}
This expected valence then drives the transition dynamics of the latent
valence state  $S_i^t\in\{\mathrm{Neg,Neu,Pos}\}$ with ordinal map $s(\mathrm{Neg})=-1$, $s(\mathrm{Neu})=0$, $s(\mathrm{Pos})=1$. Three behavioral factors shape the dynamics:
(i) baseline biases $\omega_{k\ell}$ (how likely each shift is by default),
(ii) sensitivity $\eta_k$ (how strongly expected valence pushes transitions),
and (iii) penalty $\rho_k$ (how costly big jumps are, e.g.\ from negative to positive in one step).
We parameterize transitions by expected valence, with stickiness and large–jump penalty:
\begin{equation}
\label{eq:TS}
T_{S,i}^{t}(k\to \ell)
\;=\;
\frac{\exp\!\big\{\, \omega_{k\ell} \;+\; \eta_k\,\mathbb{E}(V_i^t)\,s(\ell) \;-\; \rho_k\,|s(\ell)-s(k)| \,\big\}}
{\sum_{j}\exp\!\big\{\, \omega_{kj} \;+\; \eta_k\,\mathbb{E}(V_i^t)\,s(j) \;-\; \rho_k\,|s(j)-s(k)| \,\big\}},
\end{equation}
with baselines $\omega_{k\ell}$, sensitivity $\eta_k>0$, and penalty $\rho_k\ge 0$. 
\paragraph{Filtering and readout.}
The HMM updates the probability of each valence level,
using both the previous distribution and the transition dynamics. This yields
a smooth trajectory of affective states. The belief state over valence evolves as
\[
\gamma_{S,i}^t \propto (T_{S,i}^t)^\top \gamma_{S,i}^{t-1},
\qquad \sum_{\ell} \gamma_{S,i}^t(\ell) = 1.
\]

We can obtain a scalar valence readout in two ways:
\begin{equation}
\label{eq:valence-readout}
v_i^t = \nu^\top \gamma_{S,i}^t, \qquad \nu = (-1, 0, 1)^\top,
\end{equation}
 Here, valence is read out as a weighted average across
state probabilities, yielding a single continuous score in $[-1,1]$.

\subsection{(Expected) Valence-guided Policy selection}
We define a valence–compatible prior over policies (Eq.~\ref{eq:policy}) with inverse temperature $\lambda$ controlling decisiveness. The compatibility
function
\begin{equation}
g(E(V_i^t),\pi)=\mathbb{E}_{P(o^t\mid \pi)}\Big[\log P\big(o^t \,\big|\,
\mathbb{E}(V_i^t)\big)\Big]
\end{equation}
scores each candidate policy by how well its expected outcomes align with the
agent’s current valence state. In other words, policies are preferred if they
make the world more consistent with the agent’s affective expectations.

\subsection{(Realized) Valence-guided Model Updating}
\label{app:valence-realized}
After enacting the policy, agents would get an actual outcome of the chosen policy and have an actual emotional response (realized valence (Eq.~\ref{eq:valence-realized-app})). Conceptually, this provides a feedback signal for updating. 

When outcomes under policy $\pi_i^t$ match expectations, 
realized valence is high, reinforcing both the policy and the current self-model. 
When outcomes diverge, realized valence is low (or negative), signaling a 
need to adjust the model’s parameters, update memory with a new impact tag, 
and reconsider future policies.
 This update propagates across all latent states and parameters:

\begin{itemize}
  \item \text{Self-state posteriors:} priors $P_i^t$ are corrected to posteriors 
        $\hat P_i^t$ using the new observation $(o_i^t,o_j^t)$.
  \item \text{Valence HMM:} beliefs $\gamma_{S,i}^t$ are updated using
        expected valence $\mathbb{E}(V_i^t)$ as input, ensuring temporal
        persistence of affect.
  \item \text{Policy posterior:} the distribution over policies
        $P(\pi_i^t \mid V_i^t)$ is recalculated given the new affective state.
  \item \text{Impact encoding:} the realized impact $I^t$ is computed via
        Eqs.~\eqref{eq:impact}–\eqref{eq:tv}, summarizing prediction error
        and policy reorganization.
  \item \text{Memory write:} the tuple $(z_i^t, I^t, t)$ is appended to the
        autobiographical store $\mathcal{LTM}_i$, allowing subsequent
        self-relevance estimates to remain personalized and history-dependent.
\end{itemize}

 This way, realized valence serves as a model-updating drive that keeps affective inference aligned with lived experience.

\subsection{Table.1}
\begin{table}[H]
\centering
\begin{tabularx}{\linewidth}{@{} l l X @{}}
\toprule
\textbf{Symbol} & \textbf{Type} & \textbf{Definition / Role} \\
\midrule
$M_i$ & model & Agent $i$'s generative model (self + simulated other). \\
$o_i^t,o_j^t$ & obs & Observations at step $t$ (self / partner). \\
$P_i^{t},\hat P_i^{t}$ & state & Self-state prior / posterior after observing $(o_i^t,o_j^t)$. \\
$PE_{\text{self},i}^{t}$ & vec & Identity prediction error; we use $\|PE_{\text{self},i}^{t}\|$. \\
$a_j^t$ & act (inf.) & Inferred partner action represented in $M_i$. \\
$z_i^t$ & feat & Descriptor $\phi(o_i^t,a_j^t,P_i^t)$; $\psi(z)=z$ by default. \\
$\mathcal{LTM}_i$ & memory & Autobiographical store of $(x_k,I_k,\tau_k)$. \\
$I^t$ & impact & $\mathrm{norm}(\|PE_{\text{self},i}^{t}\|)\cdot \mathrm{TV}_\pi^{\,t}$. \\
$\mathrm{TV}_\pi^{\,t}$ & scalar & Total variation between pre/post policy distributions at $t$. \\
$S(z,x)$ & sim & Similarity kernel; $\tilde S$ is its normalization over $k$. \\
$\widetilde{SR}_t^i$ & scalar & Memory-derived self–relevance, Eq.~\eqref{eq:SR-memory}. \\
$\mathbb{E}(V_i^t)$ & scalar & Expected valence, Eq.~\eqref{eq:valence}. \\
$S_i^t,\gamma_{S,i}^{t}$ & HMM & Valence state / belief; transition $T_{S,i}^{t}$ in Eq.~\eqref{eq:TS}. \\
$\omega_{k\ell},\eta_k,\rho_k$ & params & Valence baselines, sensitivity, jump penalty. \\
$\pi_i^t,\lambda$ & policy & Action policy; inverse temperature in \eqref{eq:policy}. \\
$g(V,\pi)$ & score & Compatibility in \eqref{eq:policy}. \\
\bottomrule
\end{tabularx}
\caption{Symbols used in the simulation-ready HMM implementation.}
\end{table}
\subsection{Appendix illustration}
\begin{figure}[H]
\centering
\begin{tikzpicture}[font=\small, node distance=9mm,
  box/.style={draw, rounded corners, align=center, inner sep=4pt, fill=gray!5},
  arr/.style={-{Latex}, thick}
]
\node[box] (encode) {Encode exchange\\$z_i^{t-1}=\phi(o_i^{t-1},a_j^{t-1},P_i^{t-1})$};
\node[box, right=27mm of encode] (srmem) {Memory retrieval\\$\tilde S(z,x),I_k \Rightarrow \widetilde{SR}_{t-1}$};
\node[box, below=of encode] (pe) {Identity surprise\\$\|PE_{\text{self}}^{\,t-1}\|=\|\hat P^{t-1}-P^{t-1}\|$};
\node[box, below=of pe] (evalence) {Expected valence\\$\mathbb{E}(V^t)=\alpha\,\|PE_{\text{self}}^{\,t-1}\|\,\widetilde{SR}_{t-1}$};
\node[box, below=of evalence] (valhmm) {Valence HMM\\$T_S^t(\mathbb{E}(V^t))\Rightarrow \gamma_S^t,\, V^t$};
\node[box, below=of valhmm] (policy) {Policy selection\\$P(\pi^t\mid V^t)\propto e^{\lambda g(V^t,\pi^t)}$};
\node[box, below=of policy] (act) {Act \& update $M_i$\\observe $o^t$, update states};
\node[box, below=of act] (memory) {Write memory\\$I^t=\mathrm{norm}(\|PE_{\text{self}}^t\|)\cdot \mathrm{TV}_\pi^{\,t}$; add $(z^t,I^t,t)$};

\draw[arr] (encode) -- (pe);
\draw[arr] (srmem) |- (evalence); % memory feeds SR directly
\draw[arr] (pe) -- (evalence);
\draw[arr] (evalence) -- (valhmm);
\draw[arr] (valhmm) -- (policy);
\draw[arr] (policy) -- (act);
\draw[arr] (act) -- (memory);
\draw[arr] (memory.north) .. controls +(left:18mm) and +(left:18mm) .. (encode.west);

\end{tikzpicture}
\caption{Execution cycle for one agent at step $t$. 
Self–relevance is derived purely from autobiographical memory 
($\widetilde{SR}$), which modulates expected valence. 
The HMM handles valence dynamics, while memory ensures personalization and history dependence.}
\end{figure}

% ===========================
% Appendix B — LLNCS compatible
% Place this AFTER \appendix and AFTER Appendix A in your LLNCS source
% ===========================

\section{Curvature-Based Geometric Entropy Channel}
\label{app:curvature}

% Start equation numbering at 23 (the main text ended at 22)
\setcounter{equation}{22}

\paragraph{Goal.}
This appendix mirrors Appendix~A but focuses on the \emph{geometric} observation channel used in the main text: the edge-centric Forman--Ricci curvature (FRc) and its entropy as a compact, simulation-ready observable of dyadic network reconfiguration (rupture, repair, re-attunement).

\subsection{Setup and Notation}
Let $G_t=(V,E_t)$ be a time-varying inter-brain graph at step $t$, where $V$ indexes neural sources across the two agents and $E_t$ contains \emph{inter-brain} edges computed from a sliding window over the dual-brain data (e.g., cross-brain EEG coherence). Each edge $e_{ij}\in E_t$ carries a connectivity weight $w(e_{ij})$, while each node $v_i\in V$ may be assigned a weight $w(v_i)$ (e.g., node strength).

\begin{table}[t]
\centering
\caption{Symbols for the curvature-based entropy channel (Appendix~B).}
\label{tab:curv-notation}
\begin{tabularx}{\linewidth}{@{} l l X @{}}
\toprule
\textbf{Symbol} & \textbf{Type} & \textbf{Definition / Role} \\
\midrule
$G_t=(V,E_t)$ & graph & Inter-brain network at time $t$ (nodes $V$, edges $E_t$). \\
$w(e_{ij})$ & scalar & Edge weight for connectivity between $v_i$ and $v_j$. \\
$w(v_i)$ & scalar & Node weight (e.g., strength: sum of incident edge weights). \\
$\mathrm{FRc}(e_{ij})$ & scalar & Forman--Ricci curvature of edge $e_{ij}$ (local geometry). \\
$p(e_{ij})$ & prob. & Curvature-proportional mass: $\mathrm{FRc}(e_{ij})/\sum_{e'\in E_t}\mathrm{FRc}(e')$. \\
$H_{\mathrm{FRc}}(t)$ & scalar & Curvature entropy at $t$: $-\sum_{e\in E_t} p(e)\log p(e)$. \\
$O_{\kappa,t}$ & obs. & Geometric observation channel (we set $O_{\kappa,t}=H_{\mathrm{FRc}}(t)$). \\
$S_t$ & latent & Dyad-level hidden state (e.g., alignment / rupture). \\
\bottomrule
\end{tabularx}
\end{table}

\subsection{Forman--Ricci Curvature (FRc)}
For each edge $e_{ij}=(v_i,v_j)\in E_t$, the FRc for weighted graphs is:
\begin{equation}
\label{eq:frc}
\mathrm{FRc}(e_{ij}) \;=\; w(e_{ij})\!\left(\frac{1}{w(v_i)} + \frac{1}{w(v_j)}\right)
\;-\;
\sum_{\substack{e \sim e_{ij}\\ e\neq e_{ij}}}
\frac{\,w(e_{ij})\,}{\sqrt{\,w(e_{ij})\,w(e)\,}}\,.
\end{equation}
The first term grows when $e_{ij}$ dominates the load of its endpoints; the second term reduces curvature when many strong \emph{neighbor} edges $e\sim e_{ij}$ create parallel routes, flattening local geometry. Positive FRc typically marks edges in dense, redundant neighborhoods; strongly negative FRc highlights bridge-like edges stitching otherwise separate modules.

\subsection{Curvature-Based Entropy}
We compress the network’s geometry at time $t$ into a single volatility marker via the Shannon entropy of the FRc distribution:
\begin{align}
\label{eq:p}
p(e_{ij})
&=\;
\frac{\mathrm{FRc}(e_{ij})}{\sum_{e'\in E_t}\mathrm{FRc}(e')}
\quad\text{(curvature-proportional mass),}
\\[4pt]
\label{eq:entropy}
H_{\mathrm{FRc}}(t)
&=\;
-\sum_{e\in E_t} p(e)\,\log p(e)\,.
\end{align}
High $H_{\mathrm{FRc}}$ indicates dispersed curvature (topological disorder/volatility), while low $H_{\mathrm{FRc}}$ indicates concentration on a stable backbone (topological order). Abrupt rises in $H_{\mathrm{FRc}}(t)$ signal \emph{phase transitions} in inter-brain topology and operationalize affective rupture; subsequent drops track repair and re-attunement.

\paragraph{Practical note.}
In practice, some $\mathrm{FRc}(e)$ can be negative. To use Eq.~\eqref{eq:p}, one may (i) apply a positive shift before normalization, or (ii) work with a nonnegative transform (e.g., $|\mathrm{FRc}|$) when the analysis targets \emph{dispersion} rather than signed geometry. Either choice should be fixed \emph{a priori} and reported.

\subsection{Embedding Curvature in Active Inference}
We treat the curvature entropy $H_{\mathrm{FRc}}(t)$ as a dyad-level \emph{observation channel} $O_{\kappa,t}$ in the generative model:
\begin{equation}
\label{eq:likelihood}
P\!\big(O_{\kappa,t}\mid S_t\big)
\quad\text{with}\quad
O_{\kappa,t}\equiv H_{\mathrm{FRc}}(t)\,.
\end{equation}
Hidden states $S_t$ encode dyadic alignment (e.g., $\text{attunement}\;\leftrightarrow\;\text{low }H_{\mathrm{FRc}}$, $\text{rupture}\;\leftrightarrow\;\text{high }H_{\mathrm{FRc}}$). The agent inverts this likelihood online to infer $P(S_t\mid O_{\kappa,t},\text{other cues})$. Curvature-induced surprises (unexpected $H_{\mathrm{FRc}}$ spikes) increase the posterior probability of rupture and drive policy adaptation toward repair; sustained low $H_{\mathrm{FRc}}$ reinforces attuned policies.

\subsection{Execution Cycle (Geometry Channel)}
Figure~\ref{fig:curv-cycle} mirrors Appendix~A’s execution loop, substituting the geometric channel for valence-specific computations. Here, $O_{\kappa,t}$ continuously constrains belief updates and policy selection.

\begin{figure}[t]
\centering
\begin{tikzpicture}[font=\small, node distance=9mm,
  box/.style={draw, rounded corners, align=center, inner sep=4pt, fill=gray!5},
  arr/.style={-{Latex}, thick}
]
\node[box] (signals) {Dual-brain signals\\$\mathbf{x}_1,\mathbf{x}_2$ (window $[t\!-\!W,t]$)};
\node[box, right=24mm of signals] (net) {Inter-brain network $G_t$\\build $w(e_{ij})$ from FC};
\node[box, below=of net] (frc) {Compute FRc on $E_t$\\Eq.~(\ref{eq:frc})};
\node[box, below=of frc] (entropy) {Entropy $H_{\mathrm{FRc}}(t)$\\Eqs.~(\ref{eq:p})--(\ref{eq:entropy})};
\node[box, left=24mm of entropy] (obs) {Observation channel\\$O_{\kappa,t}=H_{\mathrm{FRc}}(t)$};
\node[box, below=of obs] (infer) {Update beliefs $P(S_t\mid O_{\kappa,t},\dots)$};
\node[box, below=of infer] (policy) {Select policy (repair/maintain)\\min.\ expected free energy};
\node[box, below=of policy] (act) {Act \& couple\\(behavior $\leftrightarrow$ coupling)};

\draw[arr] (signals) -- (net);
\draw[arr] (net) -- (frc);
\draw[arr] (frc) -- (entropy);
\draw[arr] (entropy) -- (obs);
\draw[arr] (obs) -- (infer);
\draw[arr] (infer) -- (policy);
\draw[arr] (policy) -- (act);
\draw[arr] (act.west) .. controls +(left:18mm) and +(left:18mm) .. (signals.west);
\end{tikzpicture}
\caption{\textit{Curvature-entropy observation loop (Appendix~B).} A sliding window yields $G_t$, FRc is computed per edge, and $H_{\mathrm{FRc}}(t)$ feeds an observation channel $O_{\kappa,t}$ for online inference and policy selection.}
\label{fig:curv-cycle}
\end{figure}

\newpage

\subsection{Algorithmic Skeleton (Simulation-Ready)}
\begin{algorithm}[t]
\caption{Curvature-Entropy Driven Inference Cycle (mirror of Appendix~A)}
\label{alg:curv}
\begin{algorithmic}[1]
\STATE \textbf{Inputs:} window $W$; initial $P(S_0)$; coupling/likelihood params for $P(O_{\kappa,t}\mid S_t)$; dual-brain stream or simulator.
\FOR{$t=1,2,\dots$}
  \STATE \textbf{Signals:} extract $\mathbf{x}_1[t-W:t],\,\mathbf{x}_2[t-W:t]$.
  \STATE \textbf{Network:} compute FC across brains $\Rightarrow$ weights $w(e_{ij})$; form $G_t=(V,E_t)$.
  \STATE \textbf{Curvature:} for each $e_{ij}\in E_t$, compute $\mathrm{FRc}(e_{ij})$ via Eq.~(\ref{eq:frc}).
  \STATE \textbf{Entropy:} compute $p(e)$ by Eq.~(\ref{eq:p}) and $H_{\mathrm{FRc}}(t)$ by Eq.~(\ref{eq:entropy}).
  \STATE \textbf{Observation:} set $O_{\kappa,t}\!\leftarrow\!H_{\mathrm{FRc}}(t)$; combine with other cues.
  \STATE \textbf{Inference:} update $P(S_t\mid O_{\kappa,t},\dots)$ (variational or filtering).
  \STATE \textbf{Policy:} select action/policy minimizing expected free energy (repair if rupture likely).
  \STATE \textbf{Act:} enact behavior; update coupling/state dynamics for next step.
\ENDFOR
\end{algorithmic}
\end{algorithm}

\subsection{Notes on Design Choices}
\begin{itemize}
\item \textbf{Windowing \& FC.} The window $W$ should balance temporal sensitivity and estimator stability; FC choices (e.g., correlation/coherence) should match modality and frequency-band of interest.
\item \textbf{Node weights.} $w(v_i)$ can be node strength or constant $1$ (unweighted node case); choices impact the sign/scale of FRc and should be reported.
\item \textbf{Normalization.} If signed FRc complicates Eq.~\eqref{eq:p}, adopt a fixed nonnegative mapping (shift or magnitude) chosen \emph{a priori}.
\item \textbf{Observation model.} Calibrate $P(O_{\kappa,t}\mid S_t)$ by aligning $H_{\mathrm{FRc}}$ transitions with behavioral/phenomenological markers (rupture/repair annotations) for the paradigm at hand.
\end{itemize}

\clearpage

% 

\bibliographystyle{splncs04}
\bibliography{ref}
 
\end{document}